\documentclass[12pt,twoside]{article}
\usepackage{xrb2007}
\newcommand{\integral}{{\em INTEGRAL }}
\newcommand{\xte}{{\it RXTE}}
\usepackage{epsf}
\usepackage{graphicx}
\usepackage{lscape}
\begin{document}

\session{Future Missions and Surveys}

\shortauthor{Capitanio}
\shorttitle{BeppoSAX WFC Surveys}

\title{Correlating the WFC and the IBIS hard X-ray surveys}
\author{Fiamma Capitanio}
\affil{IASF-Roma INAF. Via Fosso del Cavaliere, 100 00133 Rome Italy}
\author{Antony J. Bird}
\affil{School of Physics and Astronomy, University of Southampton, Highfield, Southampton, SO17 1BJ, UK }
\author{Memmo Federici, Angela Bazzano, Pietro Ubertini}
\affil{IASF-Roma INAF, Via Fosso del Cavaliere, 100 00133 Rome Italy}
\begin{abstract}

During the operational life of BeppoSax, the Wide Field Camera observations covered almost the full sky at different epochs. The energy coverage, wide field of view and imaging capabilities share many commonalities with IBIS, the gamma-ray telescope onboard INTEGRAL satellite.
We mosaicked all the available single pointing WFC observation  images and then we searched the map for persistent and transient sources as has been done
for the IBIS survey. This work represents the first unbiased source list compilation produced in this way from the overall WFC data set.
 
\end{abstract}

\section{Introduction}

The two Wide Field Cameras (WFCs)~\cite{Jag97} on board the BeppoSAX satellite~\cite{Boe97_1}, were mounted 180 degrees away from each other and perpendicular to the pointing direction
 of the Narrow Field Instruments (NFI) looking at two different sky zones during any NFI pointing.
 In this way, during the 6 years of  BeppoSAX operational life, the WFCs secondary mode observations covered almost all the sky, with at
least one pointing (typically 100 ksec duration).
Moreover, two times a year the WFCs observed, as the primary instrument, the galactic bulge region. The time devoted to this program corresponded 
to 8\% of the total observing time of the satellite. Twelve such campaigns were carried out amounting to a total net exposure up to 6 Msec. 

IBIS (the gamma-ray telescope on board INTEGRAL satellite) and  WFCs are both coded mask instruments, 
they have a wide field of view and an observational strategy that favours the Galactic centre zone. 
They are characterised by a complementary energy range, partially overlapped, and a good capability 
to detect transient sources thanks to their large field of view.
%
%
We analysed all the BeppoSAX Wide Field Camera pointings available providing a mosaic of the images of all the single pointing observations and 
searched the map for sources using the same softwares and techniques developed for the IBIS  survey ~\cite{Bird3}.

Each excess found in the WFCs maps was searched for in the IBIS maps in order to better distinguish between structures of 
the IBIS survey mosaics and sources.

 A comparison between the IBIS source catalogue  and the WFC one,
 especially in the overlapping energy range of the two instruments, can
 be very useful in order to better understand the nature of the sources
 detected in the IBIS survey. Besides its application to joint  
 studies with IBIS, this work is
 useful in its own right, demonstrating the effectiveness of the SAX/WFC  
 and the
 archival dataset it has produced.
%
%

\section{WFC Sky Maps Production}

We analysed all the selected OPs with the WFC Data Analysis System, extracting the images for two energy ranges: 
3-17 keV and 17-28 keV. The latter band was selected to allow a direct comparison between WFCs and IBIS.
The WFC data are organised in short observational periods (OP) of at least 100 ks. 
We selected all the OPs available in the archive of  the IASF Institute of Rome. Some of the OPs were corrupted, so that only 70\% of the total amount 
of OPs have been wed for the analysis. After the analysis, a filter has been applied in order to eliminate the noisy images, so at the end the 
final amount of data corresponds to 60\% (100 Ms) of the total observation time. The collected data 
covered all the sky although not uniformly, in fact there are zones with higher exposure such as the galactic center and the zone around the Polaris star 
(and the ``anti'' Polaris zone). This is due to the observing strategy and manoeuvre techniques of the {\it BeppoSAX} satellite. The averaged exposure is 
about $1\times 10^4$ Ks, even if there are some regions with an exposure of  two orders of magnitude less. Nevertheless these regions are quite small 
covering a zone with a radius of about 10 degrees.

\begin{figure}[!h]
\centering
\includegraphics[angle=0,scale=1.5]{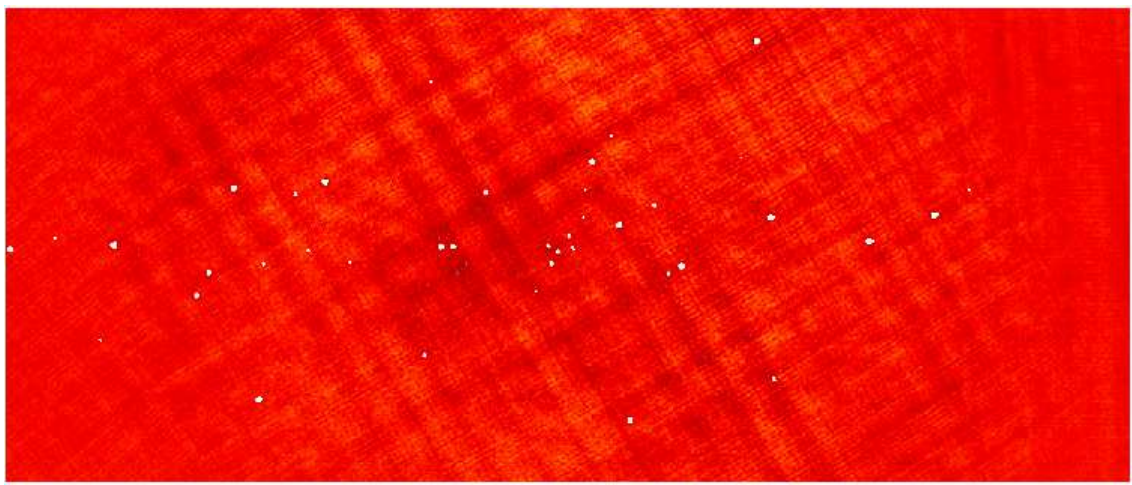}
\caption{\small{WFCs final mosaic between 3-17 keV: Zoom of the galactic centre region.}}
\label{f1map}
\end{figure}

The WFC standard software uses the IROS method (Iterative Removal of Sources) to extract sources from the shadowgrams of the WFC coded masks.
During the IROS procedures, we forced the software to extract a large number of sources for each OP image, effectively cleaning down into the noise. 
A mosaic of the images, for both energy ranges has been generated using the same software used for the IBIS survey~\cite{Bird3}. 
In this way the eventually false detections extracted by "stressing" the WFC software have been eliminated.
Two final maps have been obtained, 
Figure~\ref{f1map} shows a zoom on the galactic centre of the 3-17 keV map. 
It is clear that the analysis of WFC data carried out so far has  
concentrated on
locating transient sources, whereas our methods allow a much more  
efficient detection of weak persistent sources.

\subsection{Source List Generation}

The source positions and fluxes were identified using a barycentering method to determine the centroid of the source profile. The mean flux of the sources 
was determined from the count rate at the position of the source maximum significance, using the same software developed for the IBIS survey. 
A second method based on SExtractor 2.4.4 software~\cite{Sex} has been used to cross check the results. The list of excesses was then checked manually.
After all the checks the final list contains 236 excesses. Of these excesses 215 were identified and classified as sources while 21 do not have a
 firm identification. Of these 21 excesses, after a filtering and a visual inspection of the maps, only two have been selected as  new source candidates.
 All the WFCs excesses were compared with the \integral lowest energy range maps, 18-60 keV and 20-100 keV.

\section{Correlation between the IBIS and WFC Maps}
\begin{figure}[!h]
\centering
\includegraphics[angle=0,scale=0.7]{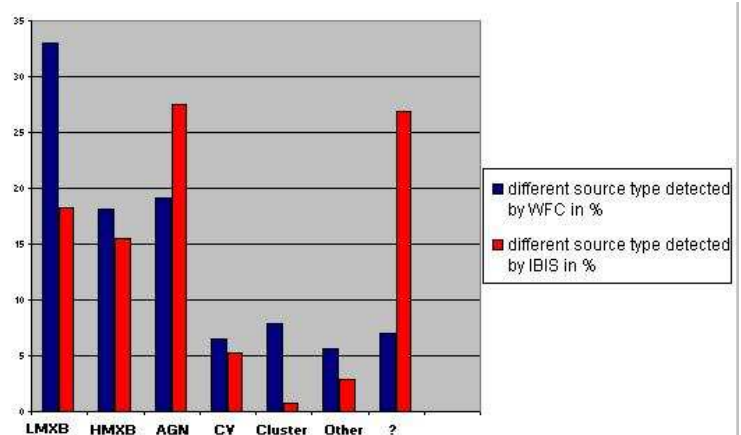}
\caption{\small{Percentage of different source types detected in WFC and IBIS/ISGRI catalogues. The question mark indicates the unidentified sources.}}
\label{WFC_IBIS_cats}
\end{figure}

The \integral and WFC source lists have been compared, both for understanding systematic effects in the 
IBIS survey and for extrapolating more information 
on the characteristics of the sources, such as  recurrency and duration of outbursts as well as spectral and time variability. We correlate the two 
excess lists obtaining two different lists of sources: the first contains all the excesses detected by both IBIS and WFCs (129 sources) while the second contains the sources detected only by WFCs.
(85 sources).
Figure~\ref{WFC_IBIS_cats} shows the different type distribution of the sources, with respect to their total number, for both IBIS and WFC catalogues.
 The distribution is different for the two instruments and this is mainly due to their different energy coverage, exposure and sensitivity: 
  as Figure~\ref{WFC_IBIS_cats} shows, the sources of the IBIS survey catalogue present a higher percentage of both AGNs and unidentified sources 
 with respect to the WFC catalogue. Otherwise the WFCs catalogue lists a higher percentage of clusters of galaxies and low-mass X-ray binaries.

\begin{figure}[!h]
\centering
\includegraphics[angle=0,scale=0.4]{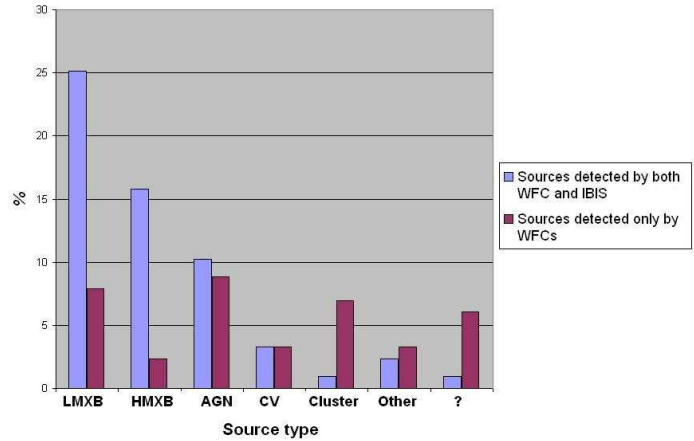}
\caption{\small{Percentage of different source types detected only by WFCs and by both WFCs and IBIS. The question mark indicates the unidentified sources.}}
\label{WFC_cor_uncor}
\end{figure}

In order to understand the reason why some sources have been detected only in the WFC mosaics, Figure \ref{WFC_cor_uncor} shows the difference in \% 
of the source types detected only by WFCs or by either IBIS and WFCs with  respect to the total of WFCs sources. Although the exposure distribution is 
similar in both instruments (both of them have a considerably higher exposure around the galactic centre zone) the diagram shows some remarkable 
differences: most of the LMXBs are present in the mosaics of both instruments.  8\% of them, detected only by WFCs, were mostly off after the \integral 
launch as can be verified from the \xte/ASM light curves, except for four LMXB that have an IBIS exposure less than 1 ks 
(X2127+119, 1H0512--401, XTE J1118+480, 4U 1700+24).

The same situation applies for the HMXBs: the ones detected only by WFCs were mostly off after the \integral launch. The number of transient sources 
in HMXB is less than the one in LMXB, this is obviously reflected in the histogram of Figure~\ref{WFC_cor_uncor}. 
The typical soft spectra of clusters of galaxies is probably the reason why this class of sources is mostly detected only by WFC.
Concerning AGN detected in the WFCs mosaics, about 40\% of them have not been observed by IBIS, and this is principally due to a low IBIS exposure. 
In fact 66\% of the AGNs detected only in the WFCs mosaics have an exposure less than 100 ks in IBIS (we extrapolated, from the AGN logN-logS published 
by Krivonos et al. 2007, that for an IBIS observation of 100 ks, in the entire IBIS FOV, a detection of $\sim$~1 AGN is expected).

\section{Comparison between our WFC catalogue and the official one.}
On July 2007 the official WFC catalogue was published \cite{Verrecchia} that is based on analysis of each single pointing observations and points mainly to the transient sources detection (a similar work, restricted to the galactic plane zone, was published in 2004 by Capitanio et al.). Our work is instead based on the searching of mosaic maps and is intended to identify persistent sources. For this reason our list of sources is slightly different from the 
one published by Verrecchia et al. 2007. The official catalogue contains 253 sources while our catalogue contains 216; while 187 sources are present in both catalogues. The 66 sources reported only by Verrecchia et al. are all transient sources, while, as we expected, the 29 found only by us are all persistent. It is important to notice that within these 29 sources there are 3 IGR sources and 2 SWIFT sources that have been discovered only after the end of the Beppo SAX mission.

\begin{figure}[!h]
\centering
\includegraphics[angle=0,scale=1]{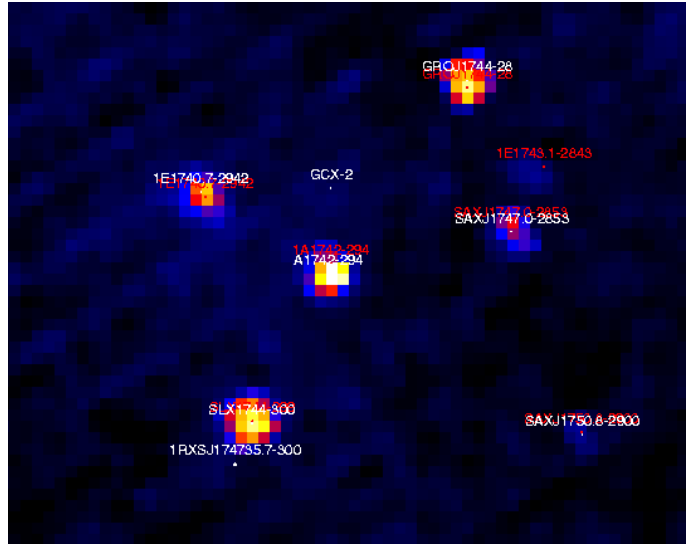}
\caption{\small{Zoom of the Galactic centre region 3-17 keV map. The sources found in the Verrecchia et al. catalogue are labelled in white colour while the sources found in our catalogue are labelled in red }}
\label{GC_res}
\end{figure}

A typical example of the different approach of our catalogue can  be  seen in Figure~\ref{GC_res} that represents a zoom of the WFC 3-17 keV map on the Galactic Centre region. The white labelled sources shown in Figure~\ref{GC_res} image are the sources reported in Verrecchia et al. 2007 catalogue, while the red ones are the sources reported in our catalogue. Most of the sources are detected by both catalogues, but for example, 1E1743.1-2843 is detected only by our catalogue. In fact 1E1743.1-2843 is a typical faint persistent source for which our work is optimised ~\cite{DelSanto}.

\section{Conclusions}
So far this work has been made with '60\% of the WFC SAX/WFC dataset'.  The work on the complete archive (95\% of the data) is still in progress. Preliminary results  shows that, by adding these new data, we get a more uniform sky exposure improving the instrument sensitivity limit and obtaining 236 confirmed source detections. 

Thanks to this work we will be able to create a combined catalogue of transient and persistent sources. Moreover, combining the WFC and INTEGRAL data, we will be able to study the long term source population characteristics.

The techniques developed for this work has been successfully applied to IBIS and WFC data. The same method could also be used with other coded mask instruments like INTEGRAL/JEM-X or SWIFT/BAT.



\begin{thebibliography}
\bibitem[Bertin \& Arnouts 1996]{Sex}
Bertin, E., \& Arnouts, S., 1996, 117, 393.
\bibitem[Boella et~al. 1997a]{Boe97_1} 
Boella, G., Butler R.C., Perola, C. et~al., 1997, A\&AS, 122, 299.
\bibitem[Bird et al. 2007]{Bird3}
Bird, A. J.,  Malizia, A., Bazzano, A., 2007, ApJS, 170, 175.
\bibitem[Capitanio et al. 2004]{cap2004}
Capitanio, F., Bazzano, A., Cocchi, M., 2004, Nuc. Phys. B. 132, 580
\bibitem[Del Santo et al. 2006]{DelSanto}
Del Santo, M.,  Sidoli, L., Bazzano, A., 2006, A\&A, 456, 1105.
\bibitem[Verrecchia et al. 2007]{Verrecchia}
Verrecchia, F., in't Zand, J. J. M., Giommi, P., A\&A 472 705 2007
\bibitem[Jager et~al. 1997]{Jag97} 
Jager, R., Mels, W.A., Brinkman, A.C. et~al., 1997, A\&AS, 125, 557.
\end{thebibliography}
\end{document}